\documentclass[twocolumn,showpacs,preprintnumbers,amsmath,amssymb]{revtex4}
\usepackage{epsfig}
\usepackage{dcolumn}% Align table columns on decimal point
\usepackage{bm}% bold math
\usepackage{graphicx}
\usepackage{color}

\begin{document}

%\preprint{preprint not for distribution}

\title{Full counting statistics of chiral Luttinger liquids with impurities}

\author{A.~Komnik$^{1}$ and H.~Saleur$^{1,2}$}

\affiliation{${}^1$~Service de Physique Th\'eorique,
CEA Saclay, F--91191 Gif-sur-Yvette, France \\
%${}^2$~Physikalisches Institut, Albert-Ludwigs-Universit\"at,
%  D--79104 Freiburg, Germany \\
${}^2$~Department of Physics, University of Southern California,
Los Angeles, California 90089-0484}

\date{\today}

\begin{abstract}
We study the statistics of charge transfer
%(full counting statistics)
through an impurity in a chiral Luttinger liquid (realized
experimentally as a quantum point contact in a fractional quantum
Hall edge state device). Taking advantage of the integrability
%of the
%corresponding Hamiltonian,
we present a procedure for obtaining
the cumulant generating function of the probability distribution
to transfer a fixed amount of charge through the constriction.
Using this approach we analyze in detail the behaviour of the
third cumulant $C_3$ as a function of applied voltage, temperature
and barrier height.
% at two values of electron interaction
%strength.
We predict that $C_3$ can be used to measure the
fractional charge at temperatures, which are several orders of
magnitude  higher than
those needed to extract the fractional charge from the measurement
of the second cumulant. Moreover, we identify the component of
$C_3$, which carries the information about the fractional charge.

\end{abstract}


\maketitle It was realized only recently that the probability
distribution function $P(Q)$ of transmitting a charge $Q$ through a
given system per unit time (also referred to as full counting statistics, FCS)
is the ultimate transport characteristics of a given device
\cite{levitovlesovik}. Along with the current voltage relation it
contains all information about the fluctuations of current and
transmitted charge.
%It turns out that
$P(Q)$ of a clean wire is
purely Gaussian, i.~e. that only the  first two cumulants (irreducible
moments) are different from zero. On the contrary, as soon as a
scatterer
%with a transmission different from unity
is inserted,
the transport is described by a binomial distribution function,
all cumulants of which are known to be nonzero
\cite{levitovlesovik,lll}. The issue of the third cumulant $C_3$
appears to be especially interesting in systems coupled to the
electrostatic environment \cite{reulet,kindnaza}. Crucial
parameters concerning the influence of the latter can be
investigated by the measurement of $C_3$.

Another question the FCS is supposed to answer is that of the
charge of current carrying excitations. The first attempt goes
back to Schottky, who already 1918 has stated that the
proportionality factor between the current autocorrelation spectra
at zero frequency and the charge current is the elementary charge
provided the transmission is weak \cite{schottky}. Using the
extrapolation of his ideas \cite{mogel1,mogel2,mogel3} to
transport phenomena in fractional quantum Hall (FQH) bars it has
been established, that the charge of current carrying excitations
is indeed a fraction of the elementary charge
\cite{dePicciotto,glattli}. Such kind of analysis is, however,
quite difficult as one has to subtract from the correlation data
the Johnson-Nyquist (JN) contribution responsible for the
equilibrium thermal fluctuations. On the other hand, it has been
suggested in \cite{reznikov} by calculations in the tunneling
regime, that the third cumulant under similar conditions may offer
much more reliable means of measuring the charge because it does
not have the JN constituent. Thus far no experimental attempt to
extract the fractional charge from higher cumulants has been
undertaken however.

The fractional charge measured in \cite{dePicciotto,glattli}
undoubtedly comes about as a result of electronic correlations.
Until now the analysis of fully interacting systems has been
restricted only to some special cases
\cite{bagrets,ourPRB,weisssaleur,AM,KT,KondoFCS,SafiSaleur}.
%perturbative approaches
%\cite{bagrets,ourPRB}, not too low temperatures \cite{bagrets},
%zero temperature \cite{weisssaleur}, and to a number of models
%which are exactly solvable at their Toulouse points
%\cite{AM,KT,KondoFCS}.
One way to make progress in understanding the FCS of fully
interacting systems is to take advantage of integrability.
Especially in 1D there are quite a number of exactly solvable
models, one notable example being the chiral Luttinger liquid
(CLL) with an impurity, which  describes in particular tunnelling
between FQH edges \cite{wen1,wen2}. In this Letter we attempt to
use the integrability of this model in order to access its FCS.

We start with the following CLL Hamiltonian in the bulk, which can
be written in terms of right/left ($R$,$L$) moving charge current
densities,
\begin{eqnarray}
  H_0 = \frac{\pi}{g} \, \int_{-\infty}^\infty \, d x \, \left( j_L^2 + j_R^2
  \right) \, ,
\end{eqnarray}
where we fix $h=1$ and where $g$ is the Luttinger liquid
interaction parameter \cite{kanefisher}. In the case of a FQH
device $g=\nu$ when $1/\nu$ is an odd integer \cite{wen1,wen2}.
The original Fermions can be rewritten as exponents of the phase
fields $j_{L,R} = \mp
\partial_x \phi_{L,R}/2 \pi$, so that the (local at $x=0$)
backscattering term (or interedge tunnelling) is given by
\begin{eqnarray}
 H_{BS} = \lambda_B \cos\left[ \phi_L(0) - \phi_R(0) \right] \, .
\end{eqnarray}
It has been shown in \cite{FLS}, that $H_0+H_{BS}$ can be mapped
(after a folding and a transformation into an even-odd basis)
onto the boundary sine-Gordon model, which is integrable by means
of Bethe ansatz (BA) technique \cite{zamolodchikov}. At $1/\nu$
odd integer -- to which we restrict now
 -- the corresponding excitations are (anti)kinks (denoted by
subscripts $\pm$), propagating in two different directions, and
$j=1 \dots 1/\nu-2$ breathers. All of them are massless and have
the dispersion relations $e_\pm = e^\theta/2$ and $e_j=\sin[ \pi
j/2(1/\nu - 1)]\, e^\theta$, where $\theta$ is referred to as
rapidity and parameterizes the particle momenta. While the
breathers are neutral, the (anti)kinks carry charge (in units of the
electron charge) $q=\mp 1$.
Furthermore, similarly to the noninteracting systems one can
introduce the level densities (per unit length and rapidity) $n_i(\theta)$
and filling fractions
$f_i(\theta)$, $i=\pm, 1, \dots 1/\nu-2$, which can be combined to
give the
density of occupied states $P_i (\theta) = n_i(\theta)
f_i(\theta)$. The filling fractions can, in turn, be rewritten in
terms of quasienergies $\epsilon_i$ according to the prescription
%\begin{eqnarray}
$f_i = P_i/n_i = 1/(1+e^{(\epsilon_i-\mu_i)/T})$,
%\end{eqnarray}
where $\mu_i$ is the chemical potential and $T$ is the
temperature. The equilibrium values of quasienergies as well as
those of density of occupied states are found by means of
thermodynamic Bethe ansatz (TBA) as derived in \cite{FLSPRL,FLS}.
The charge current $I$ can then be shown to be carried by single
(anti)kinks scattering processes across the boundary (which account
for charge transfer accross the impurity in the initial problem), breathers
only role being in establishing the equilibrium values of
$\overline{P}_\pm = \langle P_\pm \rangle$,
\begin{eqnarray}             \label{current}
 I = I(V,T_B) = \int \, d \theta \, {\cal T}(\theta) \left[
 \overline{P}_+(\theta)
 - \overline{P}_-(\theta) \right] \, ,
\end{eqnarray}
which are found using TBA \cite{FLSPRL,FLS,FS}. $V$ denotes the
voltage applied across the constriction. The quantity
 ${\cal T}(\theta)$ is the probability for a kink or anti-kink with
 rapidity $\theta$ to
 scatter off the boundary as an anti-kink or kink (so the current is
 conserved) and is given in \cite{FLS}. It strongly depends on $T_B$,
which is the energy scale generated by the backscattering strength
$\lambda_B$. The first cumulant of transmitted charge is then $C_1
= \tau I/h$ \footnote{In all cumulants we omit the trivial
prefactor $e^n$ as in accordance with our definition of $P(Q)$ we
are counting the number of elementary charges and not the actual
charge amount.}, $\tau$ being the waiting time. The zero frequency
current autocorrelation spectra (sometimes also called noise) was
obtained in \cite{FS}, the corresponding cumulant of charge given
by Eq.~(27) of \cite{FS}, supplemented with the prefactor
$\tau/h$.
%\begin{eqnarray}               \label{noise}
 %\langle \delta^2 Q \rangle
% C_2 \, h/\tau = \ell \int \int \, d \theta \, d
% \theta' \, {\cal T}(\theta) {\cal T}(\theta') \, D(\theta,
% \theta')
% \nonumber \\
% + \int \, d \theta \, {\cal T}(\theta) [1 - {\cal T}(\theta)
% ] \Big\{ \overline{P}_+(\theta) [1 - \overline{f}_-(\theta)]
% \nonumber \\
%  + \overline{P}_-(\theta) [1 - \overline{f}_+(\theta)]  \Big\} \,
%  ,
%\end{eqnarray}
%where $D(\theta, \theta')$ is an involved level occupation
%probability correlation function defined in \cite{FS}.

For  general systems, there are two ways to obtain the full $P(Q)$. One
possibility is the calculation of the cumulant generating function
$\chi(\lambda) = \sum_Q e^{i \lambda Q} P(Q)$, which can be
calculated either directly via some functional integral
\cite{nazarovlong}, or it can be reduced to a calculation of some
nonequilibrium Green's function \cite{KondoFCS,ourPRB}. The other
way is a direct evaluation of time-dependent current correlation
functions \cite{levitovlesovik} and corresponding cumulants with a
subsequent reconstruction of $P(Q)$, see e.~g. \cite{weisssaleur}.
In both cases one has to solve for nonequilibrium quantities. The
caveat of the first approach is the fact that one has to work with
some kind of nonequilibrium technique directly, e.~g. Keldysh
diagrammatics. Even  for integrable systems this cannot yet be done as
it requires the knowledge of explicitly time dependent
Green's functions, which are very difficult to extract from the
BA. In the second situation the difficulty is the causality which
expresses itself in an explicit time ordering of current
operators. It is very tempting to derive expressions which do not
contain the time ordering any more. This can indeed be done but
only for  the first two cumulants \cite{LesCh}.

Thanks to integrability however, we can rely on  yet another approach.
 Using the picture of
single particle scattering, and extending the result for non interacting
electrons \cite{lll} as has been done in the past for current and noise,
 we can {\sl identify} $\chi(\lambda)$ as the average of product of
 single integrable quasiparticle
events [(anti)kink bouncing off the boundary; we suppress
the trivial prefactor $\tau/h$ from now on],
\begin{eqnarray}            \label{guess}
 \chi(\lambda) = \Big\langle \prod_k
 \Big\{ 1 + {\cal T}(k) \Big[ (e^{i \lambda} - 1) \eta_- (1-\eta_+)
 \nonumber \\
 + (e^{-i \lambda} - 1)
  \eta_+ (1-\eta_-) \Big] \Big\} \Big\rangle \, ,
\end{eqnarray}
where the  $\eta_{\pm}$ are occupation numbers (equal to zero or
one) for  single particle kink and antikink states of momentum $k$
in  the equilibrium (and unfolded) integrable model. The brackets denote average
over the equilibrium distribution determined by the external
potential. Of course, this generating function correctly
reproduces the known noninteracting results upon
$\eta_{\pm}=\eta_{R/L}$. We claim that, by a simple extension of the arguments
used in \cite{FLS,FS}, it does give the correct result in the
integrable case as well. Note that the expression (\ref{guess}) is rather
implicit:  there are in fact  correlations between
occupation numbers, which have to be  correctly taken into account through the
BA, and the product over $k$ itself hides a complex integral, where the
density of allowed levels depends on all the filling fractions. It seems
possible to draw results for $\chi(\lambda)$ nevertheless,
but here we will contend ourselves with using it to evaluate some
cumulants.
  Defining $N_\pm = \eta_{+}(1-\eta_{-}) \pm
\eta_{-}(1-\eta_{+})$ and calculating the first three cumulants
from $C_n = (i)^n \partial^n \ln \chi(\lambda)/ \partial
\lambda^n$,
%\begin{eqnarray}            \label{currentN}
%\langle \delta^1 Q \rangle
%C_1 &=&  \sum_k {\cal T}(k)
% \langle N_-(k) \rangle \, ,
%\end{eqnarray}
%\begin{widetext}
\begin{eqnarray} \label{currentN}
&& C_1 =  \sum_k {\cal T}(k)
 \langle N_-(k) \rangle \, , \\ \label{noiseN}
 && C_2 = \sum_{k_1, k_2} {\cal T}(k_1) {\cal T}(k_2)
 \langle N_-(k_1) N_-(k_2)
 \rangle_c \nonumber \\
 &+& \sum_k {\cal T}(k) [1- {\cal T}(k)] \langle N_+(k) \rangle
 \, , \\  \label{3dCumN}
 && C_3 = \sum_k  {\cal T}(k)\left[1 - {\cal T}(k) \right]
 \left[1 - 2 {\cal T}(k)
  \right] \langle N_-(k) \rangle  \\ \nonumber
 &+& 3 \sum_{k_1, k_2}
 {\cal T}(k_1) {\cal T}(k_2) \left[1 -
 {\cal T}(k_2) \right]
 \langle
 N_-(k_1) N_+(k_2) \rangle_c
 \nonumber \\ \nonumber
 &+& \sum_{k_1, k_2, k_3} {\cal T}(k_1) {\cal T}(k_2) {\cal T}(k_3)
 \langle
 N_-(k_1) N_-(k_2) N_-(k_3) \rangle_c \, ,
\end{eqnarray}
where the subscript $c$ denotes the normal ordering. Comparing
(\ref{currentN}) to (\ref{current}) and (\ref{noiseN}) to Eq.~(27)
of \cite{FS}
%(\ref{noise})
one immediately recovers   the correct formulas for
the current and the noise, while the third
cumulant is given by the following expression, which, along with
(\ref{guess}), is the central result of the Letter,
\begin{eqnarray}         \label{central}
 %\langle \delta^3 Q \rangle
 C_3 = C_C + C_I + C_N \, ,
\end{eqnarray}
where there is a \emph{correlation contribution} just as in the
case of noise,
\begin{eqnarray}         \label{correlation}
 C_C = \ell^2 \int \int \int d \theta d \theta' d \theta''
 {\cal T}(\theta) {\cal T}(\theta') {\cal T}(\theta'')
 \nonumber \\ \times \,
 D(\theta,
 \theta', \theta'') \, ,
\end{eqnarray}
with the correlation function $D(\theta,\theta',\theta'') =
\langle \Delta(P_+ - P_-)(\theta) \Delta(P_+ - P_-)(\theta')
\Delta (P_+ - P_-) (\theta'') \rangle$ (and $\Delta P\equiv
P-\overline{P}$). The second ingredient is
the \emph{current like contribution},
\begin{eqnarray}
 C_I = \int \, d \theta \, {\cal T}(\theta)\left[1 - {\cal T}(\theta) \right]
 \left[1 - 2 {\cal T}(\theta)
  \right]
  \nonumber \\ \times
  [ \overline{P}_+(\theta) - \overline{P}_-(\theta) ] \, ,
\end{eqnarray}
which indeed resembles the current (\ref{current}) up to the
modified transmission coefficient. Finally there is a component
which has the look of the second part in the noise, and which we shall
call the \emph{noise like contribution},
\begin{eqnarray}                 \label{noiselike}
 C_N = 3 \ell \int \int d
 \theta d \theta' {\cal T}(\theta) {\cal T}(\theta') [ 1 - {\cal T}(\theta') ]
 \\ \nonumber \times
 \left\langle \left[ P_+(\theta) - P_-(\theta) \right]
 %\right. \nonumber \\   \left.
 \left\{
 P_+(\theta') [1-f_-(\theta')]
\right.  \right. \\ \nonumber + \left. \left.
  P_-(\theta') [1 -
 f_+(\theta') ] \right\} \right\rangle \, .
\end{eqnarray}
%\end{widetext}
In all these formulas, $\ell$ is the length of the system (taken to
infinity at the end), and brackets denote averages with respect to
the equilibrium distribution in the presence of chemical potentials
as determined by the TBA equations.

The main problem now is  to evaluate, using the TBA, the correlation
functions in these expressions, and integrate over them. This is an a
priori difficult endeavour, which is made possible by the use of
several tricks generalizing
 ideas used first in
\cite{FS}. The easiest part is the current
like contribution, which only requires the expectation values of
level occupation probabilities $\overline{P}_\pm$: these come out from
the numerical solution of the TBA equations.
The correlation component $C_C$ can be
extracted  by an extension of the
approach used in \cite{FS}  to evaluate
$D(\theta,\theta')$. One first defines new chemical potentials,
%\begin{eqnarray}                    \label{chempot}
$\mu_+ (\theta) = - \mu_- (\theta) = V/2T + x {\cal T}(\theta)$ .
%\end{eqnarray}
Solving the BA equations for quasienergies and calculating the
free energy per unit length  $\overline{F}$ one recovers $C_C$ by a triple
derivative according to
\begin{eqnarray}
 C_C = - \frac{1}{T} \, \frac{\partial^3 \overline{F}}{\partial x^3}
 \Big|_{x=0} \, .
\end{eqnarray}
The background of this formula is the expression for the system's
partition function as a functional integral over the level
occupation probability, see the equation between Eqs.~(5) and (6)
of \cite{FS},
%\begin{eqnarray}           \label{StatSumma}
%  Z &=& \int \left[ \prod_i {\cal D} P_i(\theta) \right] e^{- F_0
%  L/T + L \sum_i \int d \theta \mu_i (\theta) P_i(\theta) }
%  \nonumber \\
%  &=& e^{-\overline{F}L/T} \,
%  ,
%\end{eqnarray}
where the chemical potentials can be explicitly rapidity
dependent.
%as proposed in (\ref{chempot}).
%$\overline{F}$ is then
%the saddle point of the functional integral, the corresponding
%equations being the TBA equations generating the equilibrium
%values of the quasienergies \cite{FS}.
Finally, in order to calculate  the integral in the noise like part
$C_N$,  one needs to introduce a  new energy  (per unit length)
functional containing two $c$-numbers $x$ and $y$ as parameters,
\begin{eqnarray}             \label{NewE}
 E = \int d \theta \Big\{ [e_+ - a x (1-f_-)+by](\theta)
 P_+(\theta)
  \\ \nonumber
 + [e_- - a x (1-f_+)- by](\theta)
 P_-(\theta) + \sum_j e_j P_j \Big\} \, ,
\end{eqnarray}
where the very last sum is over the breathers' degrees of freedom
only and $a(\theta)$ and $b(\theta)$ are arbitrary functions of
rapidity. Modifying the energy in this way and performing the TBA
program we find following equations for the quasienergies,
\begin{eqnarray}             \label{newTBAeqs}
 && \overline{\epsilon}_i = \frac{e'_i(2)}{T} - \sum_j \int d \theta'
 \Phi_{ij} (\theta - \theta') \ln ( 1 + e^{- \overline{\epsilon}_j
 + \mu_j}) \nonumber
\\ \nonumber
 &-& \frac{a x}{T} \int d \theta' \left[ \Phi_{i+}(\theta
 - \theta') + \Phi_{i-}(\theta - \theta') \right] f_+(\theta')
 f_-(\theta') \, ,
\end{eqnarray}
with $e'_\pm(\xi) = e_\pm - a(\theta) x \pm b(\theta) y + \xi
a(\theta) x f_\mp$ for the (anti)kinks and $e'_j = e_j$ for the
breathers. The kernels $\Phi_{ij}$ can be found in
\cite{FLSPRL,FLS}. The corresponding free energy is then
\begin{widetext}
\begin{eqnarray}    \nonumber
 \overline{F} = - T \sum_i \int d \theta
  \frac{ e_i \ln \left( 1 +
 e^{-\overline{\epsilon}_i + \mu_i} \right)\left[ \overline{\epsilon}_i -
 e'_i(1)/T + (1+e^{\overline{\epsilon}_i - \mu_i}) \ln \left( 1 +
 e^{-\overline{\epsilon}_i + \mu_i} \right) \right] }{\overline{\epsilon}_i
 - e'_i(2)/T + (1+e^{\overline{\epsilon}_i - \mu_i}) \ln \left( 1 +
 e^{-\overline{\epsilon}_i + \mu_i} \right) + [a(\theta) x/T] \int d \theta'
 \left[ \Phi_{i+}(\theta - \theta') + \Phi_{i-}(\theta - \theta')\right]
 f_+(\theta') f_-(\theta') } \, .
\end{eqnarray}
\end{widetext}
%where $\tilde{e}_\pm = e_\pm - a(\theta) x \pm b(\theta) y +
%a(\theta) x f_\mp$.
On the other hand, the same free energy can be
found from the system's partition function \cite{FS}. After
subsequent derivation with respect to both $x$ and $y$, setting
them to zero afterwards and fixing $a(\theta) = {\cal
T}(\theta)[1-{\cal T}(\theta)]$, $b(\theta) = {\cal T}(\theta)$ we
immediately recognize the noise like contribution
(\ref{noiselike}),
\begin{eqnarray}
C_N = 3 T \left. \frac{\partial^2 \overline{F}}{\partial x
\partial y} \right|_{x=0, y=0} \, .
\end{eqnarray}
%We note finally that the current like contribution $C_{I}$  can be
%calculated by  redefining the chemical potentials as in
%(\ref{chempot}) with ${\cal T}(\theta)[1-{\cal
%T}(\theta)][1-2{\cal T}(\theta)]$ instead of ${\cal T}(\theta)$
%and calculating the derivative with respect to $x$.

\begin{figure}
%\vspace*{1.0cm}
\epsfig{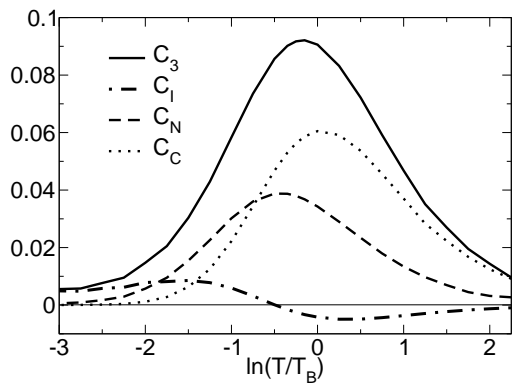}
\caption[]{\label{3dCumAsAFuncOfT} Temperature dependence of the
third cumulant and its constituents for $V/T_B=1$. }
\vspace*{0.0cm}
\end{figure}

One of the nontrivial checks of the formula (\ref{central}) is the
case $\nu=1/2$, when there are no breathers . The results of
\cite{AM,KT} are confirmed by a straightforward calculation. The
second test is the accurate reproduction of the $T=0$ result,
previously found in \cite{weisssaleur}. As can be immediately seen
from the numerical calculation of the third cumulant as a function
of temperature, Fig.~\ref{3dCumAsAFuncOfT}, all correlation
contributions vanish in the limit of low temperature. This is
quite natural as in this case all thermal fluctuations die out. The only
surviving contribution is then the current like part, which can be
computed exactly in the same way as presented in
\cite{weisssaleur}, leading, of course, to the same result.
Many other checks are possible, and the structure of $C_{3}$ is in
fact quite fascinating - we will report about this elsewhere.

Now, we would like to turn to the practical question of the
measurability  of the fractional charge through the third
cumulant. In the weak back-scattering limit at $T/V=0$, it can be
easily verified analytically \cite{FLS} that, when $V/T_{B} \gg
1$, the Schottky formula is valid $C_2/I_{BS} = \nu +
O\left[\left( T_B/V \right)^{2(1-\nu)}\right]$
%\begin{eqnarray} \nonumber
 %\lim_{V/T_B \rightarrow \infty} \frac{C_2}{I_{BS}} = \nu \left[
 %1 + \frac{a_2(\nu)}{a_1(\nu)} \left( \frac{T'_B}{V} \right)^{2(1-\nu)}
 %+ \dots \right] \, ,
%\end{eqnarray}
[here $I_{BS}$ is the backscattered current defined as
$I_{BS}=I(V,T_B=0) - I(V,T_B)$], and similarly for the third
cumulant $C_3/I_{BS} = -\nu^{2} + O\left[\left( T_B/V
\right)^{2(1-\nu)}\right]$.
%
%\begin{eqnarray}                     \label{C3TzeroLimit}
% \nonumber
% \lim_{V/T_B \rightarrow \infty} \frac{C_3}{I_{BS}} = - \nu^2
% \left[
% 1 + 3 \frac{a_2(\nu)}{a_1(\nu)} \left( \frac{T'_B}{V} \right)^{2(1-\nu)}
% + \dots \right] \, .
%\end{eqnarray}
\begin{figure}
\vspace*{1.0cm} \epsfig{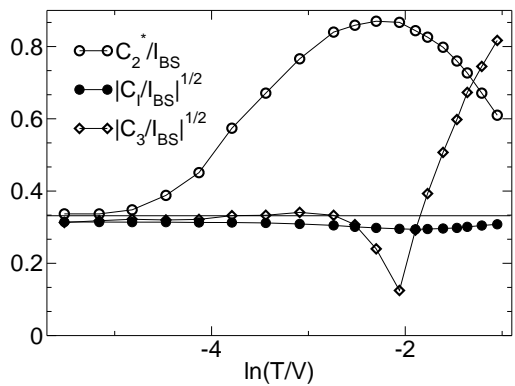}
\caption[]{\label{Robust} Temperature dependence of the square
root of the ratio of the third cumulant and of current like
contribution absolute values ($|C_3|$ and $|C_I|$) to
backscattered current $I_{BS}$ as well as of the shot noise
$C_2^{*}$, to $I_{BS}$ at $\nu=1/3$ and $T_B/V=0.1$. The cusp is
unphysical and comes about as a result of the $C_3$ sign change.}
\vspace*{0.0cm}
\end{figure}
Looking at the detailed coefficients in these expansions shows
that
 $C_2$ gives slightly  better
results for the fractional charge at $T=0$.  At finite
temperatures one has to use instead of the above $C_2$ only its
shot noise constituent  remaining after the subtraction of the
equilibrium current fluctuations: $C_{2}^{*}\equiv
C_{2}(V,T,T_{B})-C_{2}(V=0,T,T_{B})$. On the other hand, since the
equilibrium fluctuations are Gaussian, the corresponding third
cumulant is identically zero, and no subtraction is necessary for
the analysis of $C_{3}$. That circumstance led Levitov and
Reznikov to suggest that  the measurement of the bare third
cumulant would be a better way
 to access the value of  $\nu$
\cite{reznikov}. These authors also argued qualitatively that
$C_{3}$ should have a weaker dependence on temperature in the
presence of generic interactions. We would like to point out
though, that their discussion has been based on weak tunneling
results for \emph{noninteracting} particles.
%and ours, on the contrary, is a full nonperturbative verification
%of this hypothesis.

We are able to confirm this hypothesis precisely in the edge state
tunneling problem. We find for instance that for a small value of
$T_{B}/V$, while the shot noise $C_{2}^{*}$ gives a measure of the
Laughlin quasiparticles charge only in the limit $T/V$ small, the
third cumulant has essentially no dependence on temperature over a
huge range of values. This is illustrated in Fig.~\ref{Robust}.
 A similar result holds if we keep $T/V$
fixed and small, and vary $T_{B}$ (i.~e. the strength of the
backscattering potential). Interestingly, such a remarkable
behaviour of the third cumulant is entirely due to its current
like component: in Fig.~\ref{Robust} we plotted separately the
fractional charge value induced by that contribution only. It
appears to be quite insensitive to temperature change, only
slightly oscillating just below the asymptotic $\nu$.

To conclude, we have presented a procedure for obtaining the
cumulant generating function of the probability distribution
function to transfer a fixed amount of charge through a point
contact in a FQH edge state device. Using this approach we
analyzed the behaviour of the third cumulant in detail. We
predicted that the fractional charge can be much more reliably
obtained from $C_3$ than from the noise spectra and we identified
the contribution to $C_3$ which is responsible for this remarkable
property.

%This work was supported by ...?
We would like to thank H.~Grabert, A.~O.~Gogolin and B.~Trauzettel
for discussions. AK is Feodor Lynen fellow of the Alexander von
Humboldt foundation (Germany).

\bibliography{g13CumulantPaper}

\end{document}